\documentclass{jpp}
\usepackage{graphicx,bm,color}
\graphicspath{{./fig/}{./png/}}
\usepackage{amsmath}
\usepackage{amssymb}
\newcommand{\EQ}{\begin{equation}}
\newcommand{\EN}{\end{equation}}
\newcommand{\EQA}{\begin{eqnarray}}
\newcommand{\ENA}{\end{eqnarray}}
\newcommand{\eq}[1]{(\ref{#1})}

\newcommand{\Eq}[1]{equation~(\ref{#1})}
\newcommand{\Eqs}[2]{equations~(\ref{#1}) and~(\ref{#2})}

\newcommand{\App}[1]{Appendix~\ref{#1}}

\newcommand{\Sec}[1]{\S~\ref{#1}}

\newcommand{\Fig}[1]{figure~\ref{#1}}

\newcommand{\Figp}[2]{figure~\ref{#1}({#2})}
\newcommand{\Figsp}[3]{figures~\ref{#1}({#2}) and ({#3})}

\newcommand{\Figsss}[3]{figures~\ref{#1}, \ref{#2}, and \ref{#3}}
\newcommand{\Tab}[1]{table~\ref{#1}}

\newcommand{\bra}[1]{\langle #1\rangle}

\newcommand{\meanrho}{\overline{\rho}}

{}
{}

{}
{}

{}
{}
{}
{}
{}
{}
{}
{}
{}
{}
{}
{}
{}
{}

{}

{}

{}
{}
{}

\newcommand{\xx}{\bm{x}}

\newcommand{\rr}{\bm{r}}

\newcommand{\BB}{\bm{B}}

\newcommand{\AAA}{\bm{A}}

\newcommand{\uu}{\bm{u}}

\newcommand{\nab}{{\bm{\nabla}}}

\newcommand{\dd}{{\rm d} {}}

\newcommand{\const}{{\rm const}  {}}

\def\la{\mathrel{\mathchoice {\vcenter{\offinterlineskip\halign{\hfil
$\displaystyle##$\hfil\cr<\cr\sim\cr}}}
{\vcenter{\offinterlineskip\halign{\hfil$\textstyle##$\hfil\cr<\cr\sim\cr}}}
{\vcenter{\offinterlineskip\halign{\hfil$\scriptstyle##$\hfil\cr<\cr\sim\cr}}}
{\vcenter{\offinterlineskip\halign{\hfil$\scriptscriptstyle##$\hfil\cr<\cr\sim\cr}}}}}

\def\Sp{\mbox{\rm Sp}}

\def\Pm{\mbox{\rm Pr}_{\rm M}}

\def\EEM{{\cal E}_{\rm M}}

\def\EM{E_{\rm M}}

\def\cs{c_{\rm s}}

\def\xiM{\xi_{\rm M}}

\def\vA{v_{\rm A}}

\def\vAz{v_{\rm A0}}

\def\EM{E_{\rm M}}

\def\Brms{B_{\rm rms}}

\def\blue{\textcolor{blue}}
\def\blue{\textcolor{black}}

\title{Inverse cascading for initial MHD turbulence spectra
between Saffman and Batchelor}

\author{
Axel Brandenburg\aff{1,2,3,4}\corresp{\email{brandenb@nordita.org}},
Ramkishor Sharma\aff{1,2}
\and
Tanmay Vachaspati\aff{5,6}
}

\affiliation{
\aff{1}Nordita, KTH Royal Institute of Technology and Stockholm University,
Hannes Alfv\'ens v\"ag 12, SE-10691 Stockholm, Sweden
\aff{2}The Oskar Klein Centre, Department of Astronomy,
Stockholm University, AlbaNova, SE-10691 Stockholm, Sweden
\aff{3}McWilliams Center for Cosmology \& Department of Physics,
Carnegie Mellon University, Pittsburgh, PA 15213, USA
\aff{4}School of Natural Sciences and Medicine, Ilia State University,
3-5 Cholokashvili Avenue, 0194 Tbilisi, Georgia
\aff{5}Physics Department, Arizona State University, Tempe, AZ 85287, USA
\aff{6}Universit\'e de Gen\`eve, D\'epartement de Physique Th\'eorique and Centre for Astroparticle
Physics, 24 quai Ernest-Ansermet, CH-1211 Gen\`eve 4, Switzerland
}

\date{\today}

\begin{document}

\maketitle

\begin{abstract}
In decaying magnetohydrodynamic (MHD) turbulence with a strong magnetic
field, the spectral magnetic energy density is known to increase with
time at small wavenumbers $k$, provided the spectrum at low $k$ is
sufficiently steep.
This process is called inverse cascading and occurs for an initial
Batchelor spectrum, where the magnetic energy per linear wavenumber
interval increases like $k^4$.
For an initial Saffman spectrum that is proportional to $k^2$, however,
inverse cascading has not been found in the past.
We study here the case of an intermediate $k^3$ spectrum, which may be
relevant for magnetogenesis in the early Universe during the electroweak
epoch.
This case is not well understood in view of the standard Taylor expansion
of the magnetic energy spectrum for small $k$.
Using high resolution MHD simulations, we show that also in this
case there is inverse cascading with a strength just as expected from
the conservation of the Hosking integral, which governs the decay of an
initial Batchelor spectrum.
Even for shallower $k^\alpha$ spectra with spectral index $\alpha>3/2$,
our simulations suggest a spectral increase at small $k$ with time $t$
proportional to $t^{4\alpha/9-2/3}$.
The critical spectral index of $\alpha=3/2$ is related to the slope of
the spectral envelope in the Hosking phenomenology.
Our simulations with $2048^3$ mesh points now suggest inverse cascading
even for an initial Saffman spectrum.
\end{abstract}


\section{Introduction}
Standard hydrodynamic turbulence exhibits forward cascading whereby
kinetic energy cascades from large scales (small wavenumbers) to smaller
scales (larger wavenumbers) \citep{Kol41}.
This also happens in decaying turbulence, except that the rate of energy
transfer to smaller scales is here decreasing with time \citep{Bat53,Saf67}.
In magnetohydrodynamic (MHD) turbulence, the situation is in many ways
rather different.
This is primarily owing to magnetic helicity \citep{HS21}, which is a
conserved quantity in the absence of magnetic diffusivity \citep{Wol58}.
Magnetic helicity is an important property of MHD turbulence that is
not shared with hydrodynamic turbulence, even though there is kinetic
helicity that is also an invariant if viscosity is strictly vanishing
\citep{Mof69}.
However, this is no longer true when the viscosity is finite \citep{MG82}.
This is because kinetic helicity dissipation occurs faster than kinetic
energy dissipation, whereas magnetic helicity dissipation occurs more
slowly than magnetic energy dissipation for finite magnetic diffusivity
\citep{Berger84}.

The importance of magnetic helicity conservation has been recognized
long ago by \cite{Frisch+75} and \cite{PFL76} in cases when it is finite
on average.
In that case, it leads to the phenomenon of an inverse cascade.
In forced turbulence, this means that part of the injected energy gets
transferred to progressively larger scales \citep{Bra01}.
This process is at the heart of large-scale dynamos, which can be
described by what is known as the $\alpha$ effect \citep{SKR66}, and
is relevant for explaining the large-scale magnetic fields in stars and
galaxies \citep{Par79}.
In decaying MHD turbulence, on the other hand, inverse cascading leads
to a temporal increase of the magnetic energy at the smallest wavenumbers.
A similar phenomenon has never been seen in hydrodynamic turbulence,
where the spectrum at small $k$ remains unchanged.

Even when the magnetic helicity vanishes on average, there can still be
inverse cascading.
In that case, it is no longer the mean magnetic helicity density, whose
conservation is important, but the magnetic helicity correlation integral,
also known as the Hosking integral \citep{HS21, Scheko22, Zhou+22}.
In nonhelical turbulence, the possibility of inverse cascading with an
increase of spectral magnetic energy at small wavenumbers was originally
only seen for steep initial magnetic energy spectra, $\EM(k)\propto k^4$.
Here, $\EM(k)$ is defined as the spectral magnetic energy
per linear wavenumber interval and is normalized such that
$\int\EM(k,t)\,\dd k=\bra{\BB^2}/2\equiv\EEM(t)$
is the mean magnetic energy density.
Those $k^4$ spectra where motivated by causality arguments, requiring
that magnetic field correlation functions strictly vanish outside the
light cone \citep{DC03}.
Such a field can be realized by a random vector potential that is
$\delta$-correlated in space, i.e., the values of any two neighboring
mesh points are completely uncorrelated.
The magnetic vector potential $\AAA$ therefore has a $k^2$ spectrum, which
implies that the magnetic field $\BB=\nab\times\AAA$ has a $k^4$ spectrum.

For the case of a shallower $\EM(k)\propto k^2$ spectrum, no inverse
cascading has been found \citep{Reppin+Banerjee17, B+17}.
This was explained by the conservation of the magnetic Saffman integral
\citep{BL23}, which constitutes the coefficient in the leading quadratic
term of the Taylor expansion of the magnetic energy spectrum for
small $k$.

The intermediate case of a $k^3$ spectrum may be realized during the
electroweak epoch in cosmology due to a distribution of magnetic charges
as shown in \cite{Vachaspati:2020blt} and \cite{Patel:2021iik}.
The evolution of the magnetic field in this case is less clear.
\cite{Reppin+Banerjee17} reported weak inverse cascading, but it is not
obvious whether this agrees with what should be expected based on the
conservation of the Hosking integral, or whether it is some intermediate
case in which the possible conservation of both the magnetic Saffman
integral and also the Hosking integral can play a role.
Investigating this in more detail is the purpose of the present work. 

\section{Preliminary considerations}

\subsection{Relevant integral quantities in MHD}

Three important integrals have been discussed in the context of decaying
MHD turbulence.
The first two are the magnetic Saffman and magnetic Loitsyansky integrals
\citep{HS21},
\begin{eqnarray}\label{saffman}
I_{\rm SM}&=&\int\bra{\BB(\xx)\cdot\BB(\xx+\rr)}\,\dd^3\rr,\\
\label{magneticli}
I_{\rm LM}=&-&\int\bra{\BB(\xx)\cdot\BB(\xx+\rr)}\,r^2\,\dd^3\rr,
\end{eqnarray}
respectively.
Here, angle brackets denote ensemble averages, which we approximate
by volume averages.
The integrals \eq{saffman} and \eq{magneticli} are analogous to those
in hydrodynamics, but with $\BB$ being replaced by the velocity $\uu$.
The third relevant quantity is the Hosking integral
\citep{HS21, Scheko22},
\begin{equation}
I_{\rm H}=\int\bra{h(\xx)h(\xx+\rr)}\,\dd^3\rr,
\label{Hintegral}
\end{equation}
where $h=\AAA\cdot\BB$ is the magnetic helicity density.
By defining the longitudinal correlation function $M_{\rm L}(r)$ through
\EQ
\langle \BB(\xx) \cdot \BB(\xx+\rr)\rangle=\frac{1}{r^2} \frac{d}{dr}(r^3 M_{\rm L}),
\label{BBcorr}
\EN
the integrals $I_{\rm SM}$ and $I_{\rm LM}$ emerge as the coefficients
of the Taylor expansion of the magnetic energy \citep{kandu2019}.
A similar expansion also applies to the magnetic helicity variance
spectra \citep{HS21}.

For power spectra that decay sufficiently rapidly, a Taylor expansion
of $(kr)^{-1}\sin kr$ gives,
\begin{eqnarray}
\left.\Sp(\BB)\right|_{k\to0}
&=&\frac{2 k^2}{\pi} \int \frac{d}{dr}(r^3 M_{\rm L}) \left(1-\frac{k^2 r^2}{6}+...\right) dr \equiv \frac{I_{\rm SM}}{2\pi^2}k^2
+\frac{I_{\rm LM}}{12\pi^2}k^4+...,
\label{ExpSpB}\\
\left.\Sp(h)\right|_{k\to0}
&=&\frac{I_{\rm H}}{2\pi^2}k^2
+...\;.
\label{ExpSph}
\end{eqnarray}
Here, $\Sp(h)=(k^2/8\pi^3L^3)\oint_{4\pi}|\tilde{h}|^2\,\dd\Omega_k$
is the shell-integrated spectrum in volume $L^3$, the tilde marks a quantity
in Fourier space, and $\Omega_k$ is the solid angle in Fourier space, so that
$\int\Sp(h)\,\dd k=\bra{h^2}$, and likewise for $\int\Sp(\BB)\,\dd k=\bra{\BB^2}$.
The definition of shell integration implies that Parseval's theorem in the form
$\bra{h^2}L^3=\int|\tilde{h}|^2\,\dd^3k/(2\pi)^3$ is obeyed.
The magnetic energy spectrum is defined as $\EM(k,t)=\Sp(\BB)/2\mu_0$,
where $\mu_0$ is the magnetic permeability, but in the following, we
measure $\BB$ in units where $\mu_0$ is set to unity.

According to \Eq{ExpSpB}, $\Sp(\BB)$ seems to be constrained
to having only even powers of $k$ in the limit $k \to 0$.
Furthermore, $\Sp(\BB) \propto k^2$ when $I_{\rm SM}$ is finite and dominant,
and likewise, $\Sp(\BB) \propto k^4$ when $I_{\rm LM}$ is finite and dominant.
The expansion in powers of $k$ in \eqref{ExpSpB} holds, however, only
when the coefficients in the expansion are finite.
This is the case if, for example, $M_{\rm L}$ is an exponentially decaying
function of $r$.
If, on the other hand, $M_{\rm L}$ decays only as a power law, the expansion
does not hold since higher order coefficients will be divergent.
In such cases the leading order behavior in $k$ may consist of odd 
(or even arbitrary) powers of $k$.
A simple counterexample to the expansion in 
\eqref{ExpSpB}
is provided by
considering the case $r^3M_L \propto r$ for large $r$ in \eqref{BBcorr}.
The specific case of $\Sp(\BB)\propto k^3$ occurs for magnetic fields
produced during electroweak symmetry breaking as discussed in
\cite{Vachaspati:2020blt} and \cite{Patel:2021iik}.

\subsection{Competition between $I_{\rm SM}$ and $I_{\rm H}$}

Using the Taylor expansion of the magnetic energy spectrum in \Eq{ExpSpB},
we see that for initial Saffman scaling ($\EM \propto k^\alpha$
with $\alpha=2$), the magnetic Saffman integral $I_{\rm SM}$ must be
non-vanishing.
For initial Batchelor scaling ($\alpha=4$), on the other hand,
$I_{\rm SM}$ vanishes initially and cannot play a role. 
In that case, the conservation of $I_{\rm H}$ becomes important and leads
to inverse cascading, which then also implies the non-conservation of
$I_{\rm SM}$ \citep{HS21}.

Our question here is what happens in the intermediate case when
$\alpha=3$, a case already discussed in the supplemental material of \cite{HS22}.
In that situation, $\Sp(\BB)$ and $\Sp(h)$ cannot be Taylor expanded
and it is unclear whether there is then inverse cascading,
because it would require violation of the conservation of $I_{\rm SM}$,
or whether $I_{\rm SM}$ is conserved, as for $\alpha=2$, and there is
no inverse cascading.

\subsection{Growth of spectral energy at small wavenumbers}

We now want to quantify the growth of spectral energy at small
wavenumbers.
As in \cite{BK17}, we use self-similarity, i.e., the assumption that
the magnetic energy spectra at different times can be collapsed on top
of each other by suitable rescaling.
Thus, we write
\begin{equation}
\EM(k,t)=\xiM^{-\beta}\phi(\xiM k),
\label{Compensated}
\end{equation}
where $\xiM(t)=\int k^{-1}\EM(k)\,\dd k/\EEM$ is the integral scale and
$\beta$ depends on the relevant conservation law: $\beta=2$ for Saffman
scaling and $\beta=3/2$ for Hosking scaling.
This follows from the dimensions of the conserved quantity; see
\cite{BL23} for details.
Next, we assume a certain initial subinertial range scaling,
$\EM(k,0)=a_{\alpha0} k^\alpha$, where $a_{\alpha0}$ is a coefficient
determining the initial field strength.
Thus, for $k\xiM\ll1$, we can write $\phi=a_{\alpha0}(\xiM k)^\alpha$, so
\begin{equation}
\EM(k,t)=a_{\alpha}\xiM^{\alpha-\beta}\, k^\alpha
\quad\mbox{($k\xiM\ll1$)}.
\end{equation}
Assuming power-law scaling, $\xiM(t)\propto t^q$, we get 
\begin{equation}
\lim_{k\to0}\EM(k,t)\propto t^{(\alpha-\beta)\,q}.
\label{tdep_for_k0}
\end{equation}
From this, it follows that inverse cascading is possible for
$\alpha>\beta$, so $\alpha=2$ and $\beta=3/2$ could, in principle,
still give rise to inverse cascading.

Following \cite{BK17}, who assumed a selfsimilar decay, we have
$q=2/(\beta+3)$, so $q=2/5$ for $\beta=2$ and $q=4/9$ for $\beta=3/2$;
see \Tab{Tscaling} for a comparison of different theoretical possibilities
for the various exponents.
Thus, unless $I_{\rm SM}$ were conserved and there were therefore no inverse
cascading, we must expect $\lim_{k\to0}\EM(k,t)\propto t^{2/3}$ for cubic
scaling ($\EM\propto k^3$, i.e., between Saffman and Batchelor scalings)
when the Hosking integral is conserved ($\beta=3/2$ and $q=4/9$).
\blue{
Let us also note that the case $\alpha>4$ reduces to that of $\alpha=4$
after a short time; see \App{Steeper}.
In the following, however, we present numerical simulations demonstrating
that for $3/2<\alpha\leq4$, there is indeed inverse cascading with the
expected rise of spectral magnetic energy at small values of $k$.
We focus on the $\alpha=3$, but we also consider $\alpha\neq3$ to show
that inverse cascading always occurs for $\alpha>3/2$ and that the
Hosking integral is conserved.
}

%TAB1
\begin{table}\caption{
Summary of $(\alpha-\beta)\,q$ for Saffman ($\alpha=2$),
Batchelor ($\alpha=4$), and intermediate ($1.7\leq\alpha\leq3$)
spectra under the assumption that either the Saffman integral
is conserved ($\beta=2$) or the Hosking integral ($\beta=3/2$).
Two non-integer values of $\alpha$ are also considered.
\blue{
For $\alpha=6$, the subinertial range quickly becomes shallower with
time, so the value $(\alpha-\beta)\,q=2$ does not apply and is put
in parentheses.
}
}\vspace{12pt}\centerline{\begin{tabular}{lcccl}
$\alpha$ & $\beta$ & $q$ & $(\alpha-\beta)\,q$ & comment, property\\
\hline
1.7 & 3/2 & 4/9 & $4/45\approx0.09$ & non-integer scaling, assuming Hosking integral conserved\\
 2  &  2  & 2/5 &    0              & Saffman scaling, assuming Saffman integral conserved\\
 2  & 3/2 & 4/9 &  $2/9\approx0.22$ & Saffman scaling, assuming Hosking integral conserved\\
2.5 & 3/2 & 4/9 &  $4/9\approx0.44$ & non-integer scaling, assuming Hosking integral conserved\\
 3  & 3/2 & 4/9 &  $2/3\approx0.67$ & cubic scaling, assuming Hosking integral conserved\\
 4  & 3/2 & 4/9 & $10/9\approx1.11$ & Batchelor scaling, assuming Hosking integral conserved\\
 6  & 3/2 & 4/9 &   (2)             & \blue{very blue spectrum}\\
\label{Tscaling}\end{tabular}}\end{table}

\section{Simulations}

We perform simulations in a domain of size $(2\pi)^3$, so the lowest
nonvanishing wavenumber is $k\equiv k_1$.
For most runs, we use $k_1=1$, but we use $k_1=0.02$ for what we call
Runs~A and D.
For the run with $\alpha=3$ (Run~B), as well as for all other runs, we
assume that the initial magnetic energy spectrum peaks at $k_0=60\,k_1$,
and therefore we consider spectral values for $k=k_1$ to approximate
the limit $k\to0$.
We use $N^3=2048^3$ mesh points in all those simulations, so the largest
wavenumber is 1024.
This is similar to a run of \cite{Zhou+22} with $\alpha=4$, which is
here called Run~C.
We also compare with some other runs that we discuss later.
All simulations are performed with the {\sc Pencil Code} \citep{JOSS},
which solves the compressible, isothermal equations using sixth order
finite differences and a third order time stepping scheme.

In the numerical simulations, the sound speed $\cs$ is always chosen to
be unity, i.e., $\cs=1$.
The initial position of the spectral peak is at $k=k_0$ and its numerical
value is chosen to be 60 and the lowest wavenumber in the domain is
unity, or, when using the data of \cite{BL23}, $k_0=1$ and $k_1=0.02$,
so that $k_0/k_1=50$.
The magnetic diffusivity is $\eta\,k_1/\cs=2\times10^{-6}$ in Runs~B
and C, so $\eta\,k_0/\cs=1.2\times10^{-4}$.
In some runs with $\alpha=2$, we also present results for larger values
of $\eta$.
The magnetic Prandtl number, i.e., the ratio of kinematic viscosity $\nu$
to magnetic diffusivity, $\Pm=\nu/\eta$, is unity for most of our runs.
For Runs~A and D, we have $\eta\,k_0/\cs=5\times10^{-5}$
and $\nu\,k_0/\cs=2\times10^{-4}$, so $\Pm=4$.

Neither hyperviscosity nor magnetic hyperdiffusivity are being used
in any of our runs.
Hyperviscosity and magnetic hyperdiffusivity are sometimes used to
enhance the length of the inertial range.
It would give rise to different scalings, as explained in the works of
\cite{HS21} and \cite{Zhou+22}.
This led to the idea that a finite reconnection time could significantly
prolong the decay 
\citep{Zhou+2019,Zhou+2020,2021MNRAS.501.3074B}.
However, this aspect will not be pursued in the present paper.

%FIG1
\begin{figure}\begin{center}
\includegraphics[width=.7\columnwidth]{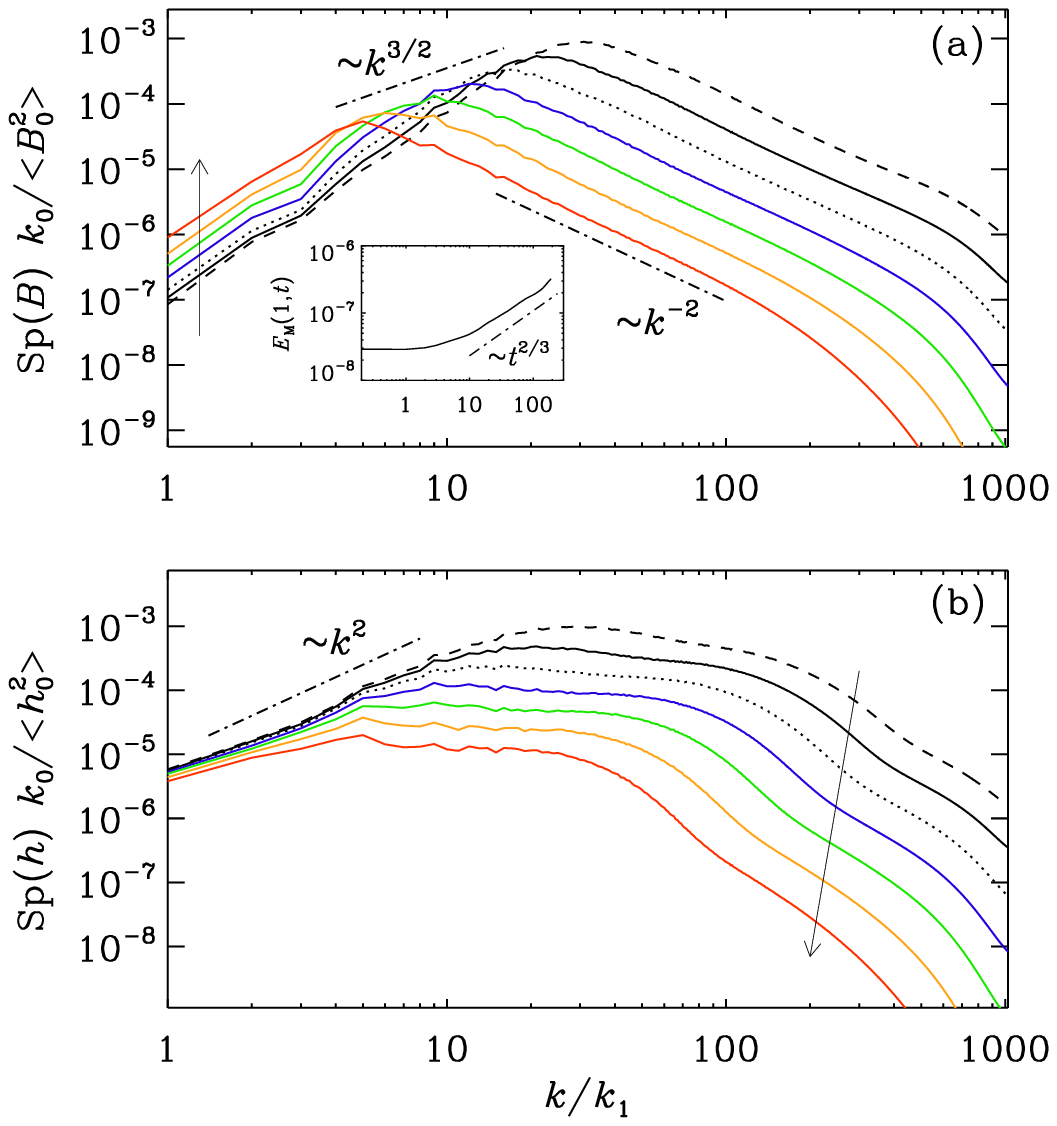}
\end{center}\caption[]{
(a) $\EM(k,t)$ and (b) $\Sp(h)$ versus $k$ for Run~B with $\alpha=3$
at $t=2$, 5, 10, 25, 50, 100, and 200.
The first three times are shown as black dashed, solid, and dotted lines.
The next four times are shown as solid blue, green, orange, and red lines.
The upward arrow in (a) indicates the direction of time.
The inset in (a) shows that $\EM(k_1,t)\propto t^{2/3}$.
}\label{rspec_select_hoskM_k60del2bc_k3}\end{figure}

The initial magnetic field strength is characterized by the Alfv\'en
speed $\vA\equiv\Brms/\sqrt{\rho}$.
For most of our runs, we choose rather strong magnetic fields with an
initial value $\vAz/\cs\approx0.87$.

\subsection{Inverse cascading}

The results for the magnetic energy and helicity variance spectra are
shown in \Fig{rspec_select_hoskM_k60del2bc_k3}, which shows inverse
cascading with $\EM(k_1,t)\propto t^{2/3}$ and $\Sp(h)=\const$ for
$k\to0$.
The temporal increase at low $k$ is compatible with \Tab{Tscaling}
for $\alpha=3$, $\beta=3/2$, $q=4/9$, and thus $(\alpha-\beta)\,q=2/3$.
\blue{
For general values $3/2\leq\alpha\leq4$, the spectral increase at small
$k$ is proportional to $t^{4\alpha/9-2/3}$.
}

%FIG2
\begin{figure}\begin{center}
\includegraphics[width=\columnwidth]{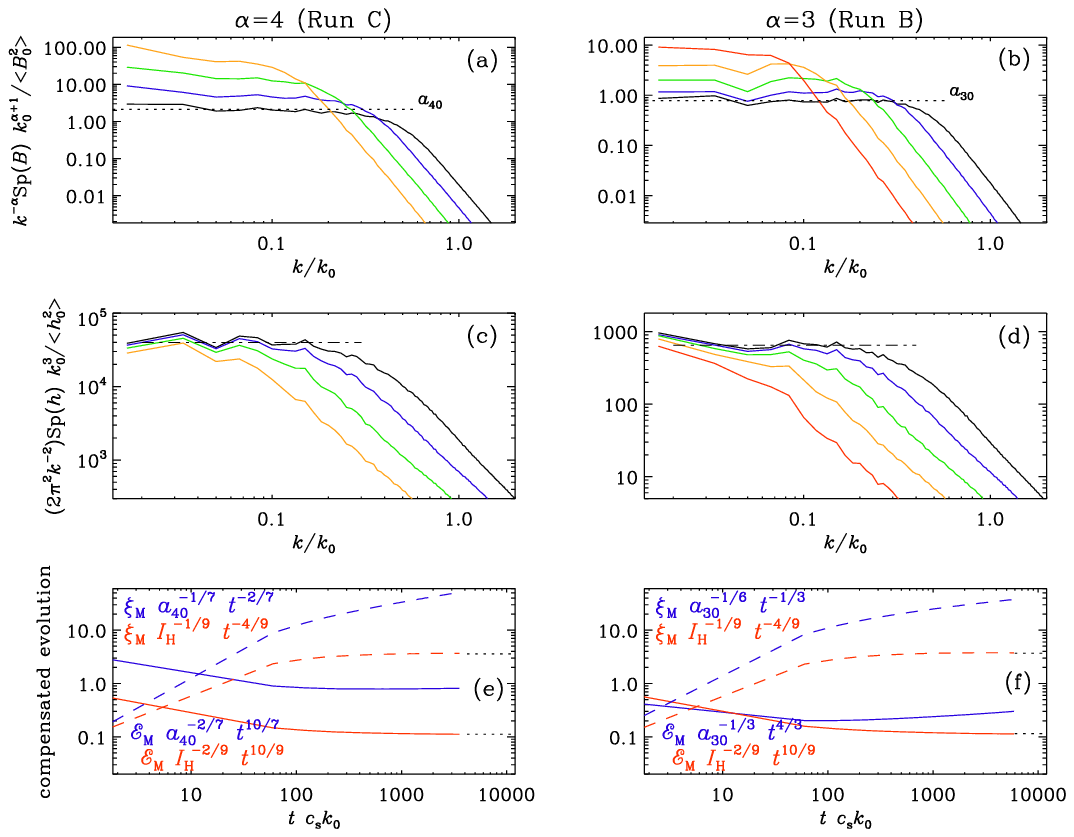}
\end{center}\caption[]{
Compensated spectra showing that $\lim_{k\to0}\Sp(\BB)/k^\alpha$
is not constant, and that instead the Hosking integral is conserved.
The left and right sides show a comparison between $\alpha=4$
(Batchelor spectrum, left) and $\alpha=3$ (right).
The times are $t=2$ (black), 6 (blue), 20 (green),
60 (orange), and 190 (red).
For $\alpha=4$ (left column), $t=190$ is not available.
In the last two panels, we see that the red lines asymptote to
constants, compatible with earlier work (BL23).
In (e) and (f), the dashed curves denote the compensated time dependences
of $\xi_{\rm M}(t)$ and the solid ones refer to the compensated
dependences of ${\cal E}_{\rm M}(t)$.
Thus, we plot $\xiM I_{\rm H}^{-1/9} t^{-4/9}$ and
$\EEM I_{\rm H}^{-2/9} t^{10/9}$,
which are non-dimensional and should approach constants.
The dotted lines mark the approximate positions of the asymptotic values
of the nondimensional constants in the Hosking scalings.}
\label{rspec_select_comp_k60del2bc_k3}\end{figure}

In the supplemental material of \cite{HS22}, it was proposed that the
evolution for $\alpha \le 3$ deviates from self-similarity at intermediate
times, and that the spectrum might show a ``pile-up'' to the left of
the peak where it would locally be approaching $k^4$ scaling.
In fact, this was already proposed by \cite{Vachaspati:2020blt}; see
his figure~16.
While we cannot exclude the possibility of a short $k^4$ range,
\Figp{rspec_select_hoskM_k60del2bc_k3}{a} suggests that such a tendency
could at best be identified at intermediate times.
However, according to the phenomenology of \cite{HS22}, this range should
become wider at later times.
This is certainly not the case in the simulations, but there is the worry
that at late times, our results become affected by finite size effects;
see the blue and green curves for $t=25$ and 50, respectively.

At early times, our simulations are poorly resolved \blue{and one might
wonder whether they are then still reliable.}
Poor resolution can be seen by the lack of a proper dissipation range
in \Figp{rspec_select_hoskM_k60del2bc_k3}{a} for $t=2$.
At later times, however, the simulations are certainly well
resolved and inverse cascading is still found to persist.
\blue{Thus, we argue that the initial phase does not adversely affect our results.
Indeed, sufficiently small viscosity and magnetic diffusivity are crucial
for being able to verify the expected Hosking scaling.}

Next, we compare in \Fig{rspec_select_comp_k60del2bc_k3} compensated
spectra, which allow us to determine $a_{\alpha}=\Sp(\BB)/2k^\alpha$
and $I_{\rm H}\to2\pi^2\Sp(h)/k^2$ for small $k$, where the $k$-dependence
of those compensated spectra is approximately flat.
The fact that the magnetic Saffman integral is not conserved is evidenced
by the increase in the height $a_\alpha(t)$ of the compensated
magnetic spectra; see \Figsp{rspec_select_comp_k60del2bc_k3}{a}{b}.
Only for $\alpha=2$ does the height stay nearly constant, as was already
verified by \cite{BL23}.
In that case, we have $I_{\rm SM}=4\pi^2 a_2$.
However, we return to this aspect in \Sec{SaffmanCase}, where our higher
resolution simulations now suggest that even in that case the decay is
governed by the conservation of the Hosking integral, rather than the
magnetic Saffman integral.

In \Fig{rspec_select_comp_k60del2bc_k3}(d), we see that $\Sp(h)$ shows a
distinctly downward trend with $k$ for the smallest $k$ values.
This suggests that the conservation property of $I_{\rm H}$ begins to
deteriorate, especially at late times.
To clarify this further, even more scale separation would be useful, i.e., a
run with a larger value of $k_0$.
Such runs at a resolution of $2048^3$ mesh points are, however, rather
expensive, but it is interesting to note that, even for the case of an
initial $k^4$ spectrum, the compensated spectra show a similar downward
trend with $k$ when the numerical resolution is only $1024^3$; see
Figure~3(d) of \cite{BL23}, which corresponds to our Run~D.
It should also be noted that in \Fig{rspec_select_comp_k60del2bc_k3}(d),
the last time is $t=190/\cs k_1$, while in
\Figp{rspec_select_comp_k60del2bc_k3}{c}, the last time is only $60$.
The two times correspond to $t\,\eta k_0^2\approx1.4$ and $0.4$,
respectively.

\subsection{Universal scaling constants}

Given that $I_{\rm H}$ is reasonably well conserved and enters the
evolution of magnetic energy and correlation length, as well as the
spectral envelope of the peak, through dimensional arguments, it is
useful to determine the nondimensional coefficients in these relations.
This was done recently for the cases $\alpha=2$ and $\alpha=4$;
see \cite{BL23}, who computed the coefficients $C_{\rm H}^{(\xi)}$,
$C_{\rm H}^{({\cal E})}$, and $C_{\rm H}^{(E)}$, defined through the relations
\begin{equation}
\xiM(t)=C_{i}^{(\xi)}\,I_{i}^{\sigma}t^q,\quad
\EEM(t)=C_{i}^{({\cal E})}\,I_{i}^{2\sigma}t^{-p},\quad
\EM(k)=C_{i}^{(E)}\,I_{i}^{(3+\beta)/\sigma}(k/k_0)^{\beta},
\label{GeneralFits}
\end{equation}
where the index $i$ on the integrals $I_i$ and the coefficients
$C_{i}^{(\xi)}$, $C_{i}^{({\cal E})}$, and $C_{i}^{({E})}$ stands for
SM or H for magnetic Saffman and Hosking scalings, respectively, and
$\sigma$ is the exponent with which length enters in $I_{i}$: $\sigma=5$
for the magnetic Saffman integral ($i={\rm SM}$) and $\sigma=9$ for the
Hosking integral ($i={\rm H}$).
In the following, we focus on the case $i={\rm H}$, but refer to
\cite{BL23} for comparisons with $i={\rm SM}$.
We recall that $k_0$ is the initial position of the spectral peak.
Note that the last expression of \Eq{GeneralFits} describes an envelope
under which $E(k,t)$ evolves; see \Figp{rspec_select_hoskM_k60del2bc_k3}{a}
for an example.

In principle, the nondimensional coefficients $C_{\rm H}^{(\xi)}$,
$C_{\rm H}^{({\cal E})}$, and $C_{\rm H}^{({E})}$ could depend on other
quantities characterizing the system, for example the magnetic Reynolds
number, but they may also be universal, just like for the Kolmogorov
constant in the kinetic energy spectrum.
To begin assessing the degree of universality of these nondimensional
coefficients, we now consider the empirical laws $\xiM(t)$, $\EEM(t)$,
and $\EM(k,t)$ for the new case of $\alpha=3$.

%TAB2
\begin{table}\caption{
Summary comparison of the coefficients in the relations for $\xiM(t)$,
$\EEM(t)$, and $\EM(k,t)$ for different values of $\alpha$.
The numbers in parentheses indicate that the slope $\beta$
is incompatible with the value of $\alpha$.
}\vspace{12pt}\centerline{\begin{tabular}{ccccccccc}
Run & $\alpha$ & $C_{\rm H}^{(\xi)}$ & $C_{\rm H}^{({\cal E})}$ & $C_{\rm H}^{(E)}$ &
$\eta k_0/\cs$ & $\EEM^2\xiM^5\,k_0^5/(\meanrho\cs^2)^2$ & $I_{\rm H}/I_{\rm H}^{\rm ref}$ & resol. \\
\hline
O &1.7& 0.14 & 3.8 & 0.038  &$7\times10^{-3}$& $1.1\times10^{-2}$ &  3.5 & $2048^3$ \\% k60del2bc_k1p7
Q & 2 & 0.13 & 3.7 & 0.038  &$7\times10^{-3}$& $1.9\times10^{-3}$ &  6.3 & $2048^3$ \\% k60del2bc_k2p0b
A & 2 & 0.15 & 3.8 & (0.04) &$5\times10^{-5}$& $1.2\times10^{-5}$ &  6.5 & $1024^3$ \\% 1024c_k2_k002
A1& 2 & 0.18 & 2.1 & (0.04) &$2\times10^{-4}$& $1.7\times10^{-5}$ &  4.3 & $ 512^3$ \\% 512c_k2_k002_2em4
A2& 2 & 0.24 & 0.8 & (0.04) &$1\times10^{-3}$& $1.7\times10^{-5}$ &  3.0 & $ 512^3$ \\% 512c_k2_k002_1em3
P &2.5& 0.12 & 3.9 & 0.038  &$7\times10^{-3}$& $2.8\times10^{-3}$ &  9.6 & $2048^3$ \\% k60del2bc_k2p5
B & 3 & 0.12 & 3.7 & 0.025  &$7\times10^{-3}$& $1.7\times10^{-3}$ & 12.2 & $2048^3$ \\% k60del2bc_k3
C & 4 & 0.11 & 3.6 & 0.037  &$7\times10^{-3}$& $9.4\times10^{-4}$ &  8.3 & $2048^3$ \\% k60del2bc
D & 4 & 0.13 & 3.5 & 0.037  &$5\times10^{-5}$& $4.5\times10^{-6}$ & 17.5 & $1024^3$ \\% 1024c_k4_k002
\label{Tcomparison}\end{tabular}}\end{table}
% /cfs/klemming/scratch/b/brandenb/axel/chiral_fluids/turbulent_decay/k60del2bc_k3
% table entries are determined with phosk2.pro

As in earlier work, the nondimensional constants in the scaling laws
for $\alpha=3$ agree with those found earlier for $\alpha=4$ \citep{BL23}.
Specifically, we have
\begin{equation}
\xiM(t)\approx0.12\,I_{\rm H}^{1/9}t^{4/9},\quad
\EEM(t)\approx3.7\,I_{\rm H}^{2/9}t^{-10/9},\quad
\EM(k,t)\la0.025\,I_{\rm H}^{1/2}(k/k_0)^{3/2}.
\label{HoskingFits}
\end{equation}
The quality of these asymptotic laws can be seen from the red lines in
\Figsp{rspec_select_comp_k60del2bc_k3}{e}{f}.
The blue lines show the case if the Saffman integral were conserved.
As explained above, those are based on the values of $a_{\alpha0}$
indicated in \Figsp{rspec_select_comp_k60del2bc_k3}{a}{b}.
A comparison of the coefficients with those found by \cite{BL23} is
given in \Tab{Tcomparison}.
Note that in \Figsp{rspec_select_comp_k60del2bc_k3}{e}{f}, the solid and dashed blue lines show an
asymptotic upward trend, reflecting again that the magnetic Saffman
integral is not conserved.

\subsection{The normalized Hosking integral}

The runs of \cite{BL23} had different mean magnetic energy densities
and also the minimum wavenumber $k_1$ was not unity, but $k_1=0.02$,
unlike the present cases, where $k_1=1$.
Instead, the peak of the initial spectrum, $k_0$, was then chosen to
be unity.
To compare such different runs, it is necessary to normalize $I_{\rm SM}$
and $I_{\rm H}$ appropriately.
On dimensional grounds, $I_{\rm SM}$ is proportional to $\EEM\xiM^3$
and $I_{\rm H}$ is proportional to $\EEM^2\xiM^5$.
By approximating the spectrum as a broken power-law, as in \cite{Zhou+22},
\EQ\label{mag_spec}
\EM(k)=\left\{\begin{aligned}
& E_\text{peak}\left(k/k_\text{peak}\right)^\alpha, & k\leq k_\text{peak}, \\
& E_\text{peak}\left(k/k_\text{peak}\right)^{-s}, & k> k_\text{peak},
\end{aligned}\right.
\EN
where $s=5/3$ and $s=2$ were used to represent the inertial range slopes
at early and late times, respectively, we find
\EQ
k_\text{peak}\xiM=\frac{\alpha^{-1}+s^{-1}}{(\alpha+1)^{-1}+(s-1)^{-1}},
\quad
E_\text{peak}=\frac{\EEM\xiM}{\alpha^{-1}+s^{-1}}.
\EN
For $\alpha=2$, we find the following reference values for the
Saffman integral:
\EQ
I_{\rm SM}^{\rm ref}=2\pi^2\times\left\{\begin{aligned}
& 250/99 \quad\mbox{(for $s=5/3$)}, \\
&  16/9  \quad\mbox{(for $s=2$)}.
\end{aligned}\right.
\EN
For other values of $\alpha$, the value of $I_{\rm SM}^{\rm ref}$ is not
meaningful and only $I_{\rm H}^{\rm ref}$ is computed for other values
of $\alpha$ by using equations (2.14) and (4.5) in \cite{Zhou+22}.
It is given by
\EQ
I_{\rm H}^{\rm ref}=8\pi^2 \EEM^2 \xiM^5
\left(\frac{(\alpha+1)^{-1}+(s-1)^{-1}}{(\alpha^{-1}+s^{-1})^{5/3}}\right)^3
\left(\frac{1}{2\alpha-3}+\frac{1}{2s+3}\right).
\EN
In calculating the above expression, we assumed the magnetic field
distribution to be Gaussian and its spectrum to be of the form as given
in \Eq{mag_spec}.
These reference values are summarized in \Tab{Tcomparison2}.

%TAB3
\begin{table}\caption{
Reference values for $I_{\rm SM}$ and $I_{\rm H}$ for different combinations
of $\alpha$ and $s$.
}\vspace{12pt}\centerline{\begin{tabular}{ccllll}
$\alpha$ & $s$ & \multicolumn{2}{c}{$I_{\rm SM}^{\rm ref}/(\EEM\xiM^3)$} & \multicolumn{2}{c}{$I_{\rm H}^{\rm ref}/(\EEM^2\xiM^5)$} \\
\hline
1.7&5/3 & \multicolumn{2}{c}{---}           & $2\pi^2\, 4437053125/151086708$                        & $\approx 579.7$  \\
1.7& 2  & \multicolumn{2}{c}{---}           &    $2\pi^2\, 90870848/5097897$                     & $\approx 351.9$  \\
 2 & 5/3 & $2\pi^2\, 250/99$ & $\approx49.8$ & $2\pi^2\, 100000/5643 $ & $\approx 349.8$  \\
 2 &  2  & $2\pi^2\,  16/9 $ & $\approx35.1$ & $2\pi^2\,   2048/189  $ & $\approx 213.9$  \\
2.5& 5/3 & \multicolumn{2}{c}{---}           & $2\pi^2\, 390625/26068$ & $\approx 295.8$  \\
2.5&  2  & \multicolumn{2}{c}{---}           & $2\pi^2\, 200000/21609$ & $\approx 182.7$  \\ %same as for alpha=3 (!)
 3 & 5/3 & \multicolumn{2}{c}{---}           & $2\pi^2\, 253125/17024$ & $\approx 293.5$  \\
 3 &  2  & \multicolumn{2}{c}{---}           & $2\pi^2\,    324/35   $ & $\approx 182.7$  \\
 4 & 5/3 & \multicolumn{2}{c}{---}           & $2\pi^2\,   5120/323  $ & $\approx 312.9$ \\
 4 &  2  & \multicolumn{2}{c}{---}           & $2\pi^2\, 131072/13125$ & $\approx 197.1$ \\
\label{Tcomparison2}\end{tabular}}\end{table}

In \Tab{Tcomparison}, we list the ratio $I_{\rm H}/I_{\rm H}^{\rm
ref}$, where $I_{\rm H}^{\rm ref}\propto\EEM^2\xiM^5$ is given in
\Tab{Tcomparison2}.
%We have used here the actual values of $\alpha=2$, 3, or 4, and $s=2$ in
We have used here the actual values of $\alpha$, and $s=2$
in all cases which describes the late time inertial range; see
\Figp{rspec_select_hoskM_k60del2bc_k3}{a}.

The former ratio, $I_{\rm H}/I_{\rm H}^{\rm ref}$, varies only little,
because the Hosking integral is always reasonably well conserved, except
when the magnetic diffusivity is large.
Near $t\eta k_0^2\approx0.1$, the ratio has for all runs a well
distinguished maximum, which is the value we quote in \Tab{Tcomparison}.
These values tend to be about 20\% larger than those at the end of the
run, which are the reference values given in \Tab{Tcomparison}.

It is interesting to note that $I_{\rm H}/I_{\rm H}^{\rm ref}$ is about
twice as large on the finer mesh (Run~C) than on the coarser mesh (Run~D).
This is somewhat surprising.
It should be noted, however, that Run~C with a larger mesh had actually
a larger magnetic diffusivity ($\eta k_0/\cs=7\times10^{-3}$) than Run~D
($\eta k_0/\cs=5\times10^{-5}$); see \Tab{Tcomparison}.
It is therefore possible that Run~D was actually underresolved and that
this was not noticed yet.

\subsection{Comments on non-Gaussianity}

\begin{figure}\begin{center}
\includegraphics[scale=0.7]{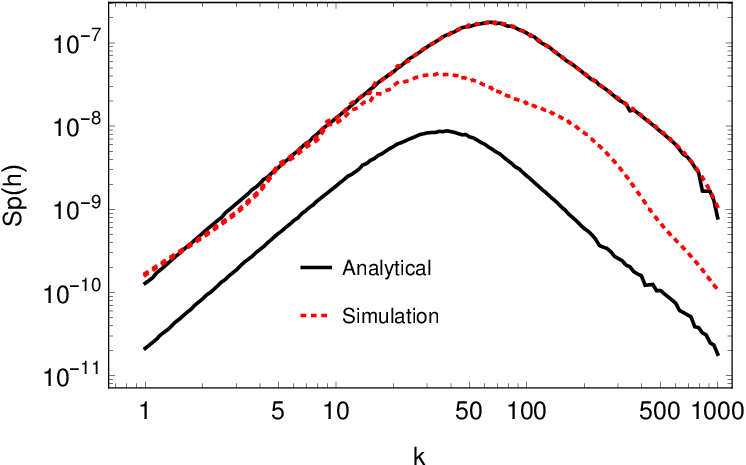}
\end{center}\caption[]{
$\Sp(h)$ at $t=0$ and $1$.
The dotted red curves represent the spectra obtained from the simulation
and the solid black curves represent the spectra calculated by assuming the
magnetic field distribution to be Gaussian.
}\label{Gaussianity_check}\end{figure}

The question of non-Gaussianity is important in many aspects of cosmology.
Not all its aspects are captured by kurtosis or skewness.
In the work of \cite{Zhou+22}, it was already pointed out that, although
the kurtosis was only slightly below the Gaussian value of three, there
was a very strong effect on the statistics of the fourth order moments
that enter in the calculation of $I_{\rm H}$ and $\Sp(h)$.
In \Fig{Gaussianity_check}, we compare $\Sp(h)$ at the initial and a later
time from the numerical calculation and the semi-analytical calculation
based on the actual magnetic energy spectra, assuming Gaussian statistics.
As in \cite{Zhou+22}, we find also here a ten-fold excess of the actual
spectra compared with the value expected based on the assumption of
Gaussianity.

%FIG3

%\subsection{How special is the Saffman scaling for $\alpha=2$?}
\subsection{Scaling for non-integer values of $\alpha$}

We now address in more detail the case $\alpha=1.7$,
for which \Eq{tdep_for_k0} with $\beta=3/2$ would predict
$\lim_{k\to0}\EM(k,t)\propto t^{4\alpha/9-2/3}=t^{4/45}\approx t^{0.09}$.
These runs are listed in \Tab{Tcomparison} as Runs~O and P.
We have seen that, for small magnetic diffusivity,
$I_{\rm H}$ is well conserved for all values of $\alpha$.
On the other hand, $I_{\rm SM}$ appears to be well
conserved in the special case of $\alpha=2$.
One possibility is therefore that, as long as $\alpha>2$, we have
inverse cascading, but not for $\alpha\leq2$.
But the argument for not expecting inverse cascading relies heavily on
the existence of $I_{\rm SM}$ and that it is non-vanishing.
If we accept that for $\alpha=3$, $\Sp(\BB)$ cannot be expanded in terms of
$k^2$ and $k^4$, then this would also be true for $\alpha=1.7$, which
is a value between $3/2$ and $2$.
One might therefore expect that also in this case, $I_{\rm SM}$ would
not be conserved, and that the decay is governed by the conservation of
$I_{\rm H}$.
This possibility was already listed \Tab{Tscaling}.

%FIG4
\begin{figure}\begin{center}
\includegraphics[width=.62\columnwidth]{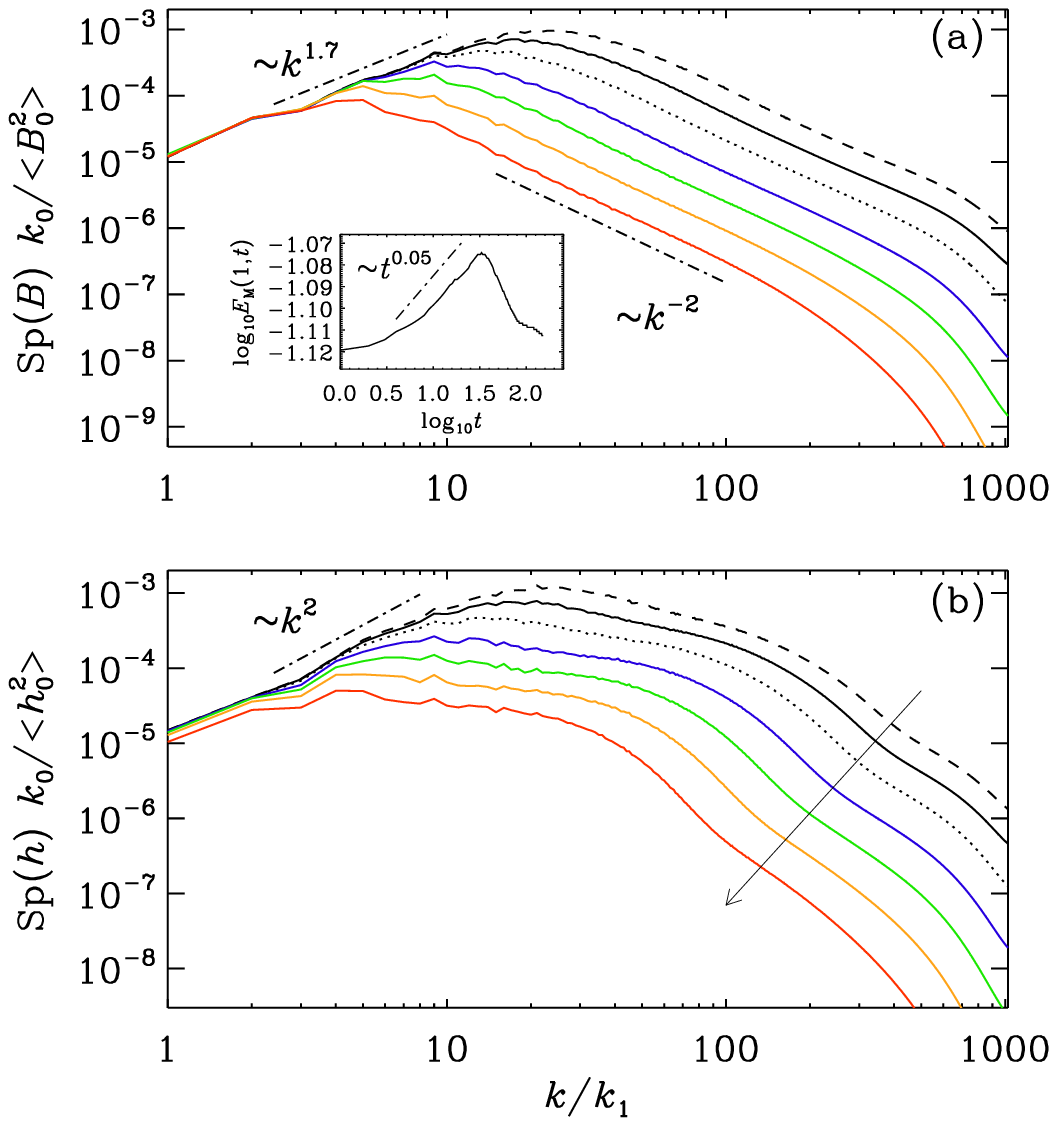}
\end{center}\caption[]{
Similar to \Fig{rspec_select_hoskM_k60del2bc_k3}, but for
$\alpha=1.7$.
}\label{rspec_select_hoskM_k60del2bc_k1p7}\end{figure}

In \Figp{rspec_select_hoskM_k60del2bc_k1p7}{a} we show that
there is no noticeable growth of $\lim_{k\to0}\EM(k,t)$.
The inset, however, does show that there is an intermediate
phase with a very weak growth $\propto t^{0.05}$.
Given that the theoretically expected growth $\propto t^{0.09}$
is already very small, and that the degree of conservation of
$I_{\rm H}$ is also limited by a finite Reynolds number, as seen in
\Figp{rspec_select_hoskM_k60del2bc_k1p7}{b} showing a premature decline of
$\Sp(h)/k^2$ in time at small $k$, it is indeed possible that at larger
resolution and smaller magnetic diffusivity, clearer inverse cascading
might emerge.

%FIG5
\begin{figure}\begin{center}
\includegraphics[width=.62\columnwidth]{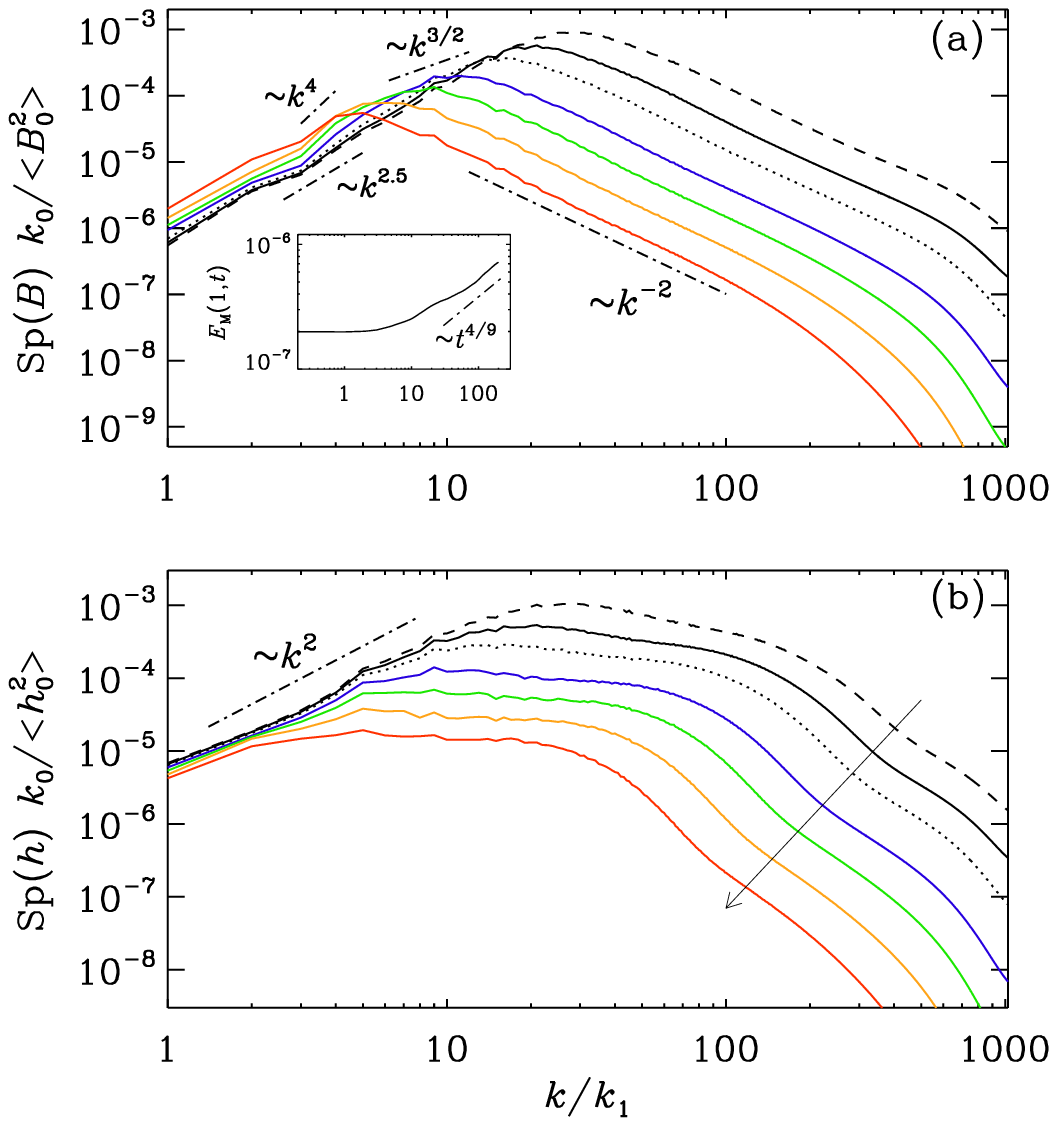}
\end{center}\caption[]{
Similar to \Fig{rspec_select_hoskM_k60del2bc_k3}, but for
$\alpha=2.5$.
Note that for the green line at $t=50$, there is some evidence for a
short range with a steeper spectrum, possibly $\propto k^4$.
}\label{rspec_select_hoskM_k60del2bc_k2p5}\end{figure}

Next, we consider the case $\alpha=2.5$.
The results are shown in \Fig{rspec_select_hoskM_k60del2bc_k2p5}.
We see inverse cascading that is compatible with
$a_{2.5}(t)\propto t^{4/9}=t^{0.44}$.
Note that for the intermediate time $t=50$, there is some evidence for
a short range with a steeper spectrum, but it would hardly be as steep
as $k^4$.

\subsection{Reassessment of the Saffman case}
\label{SaffmanCase}

Given that there is now some evidence for inverse cascading for
$\alpha=1.7$, it is reasonable to re-address earlier evidence for the
absence of inverse cascading for $\alpha=2$.
We must remember that the results of \cite{BL23} for $\alpha=2$ were
obtained at a resolution of $1024^3$ mesh points using a value of the
magnetic Reynolds number that was possibly too large for that resolution.
\blue{
More importantly, however, a superficial inspection of the spectral
evolution may not suffice.
}
We have therefore repeated such a calculation using otherwise the
same parameters as in \Figsss{rspec_select_hoskM_k60del2bc_k3}
{rspec_select_hoskM_k60del2bc_k1p7}{rspec_select_hoskM_k60del2bc_k2p5}
and compared the evolution of the spectral magnetic energy at low $k$
with that expected theoretically.
\blue{
Our initial result suggested that a larger scale separation would be
needed to obtain reliable results; see \App{FiniteSizeEffects}.
}

%FIG6
\begin{figure}\begin{center}
\includegraphics[width=.62\columnwidth]{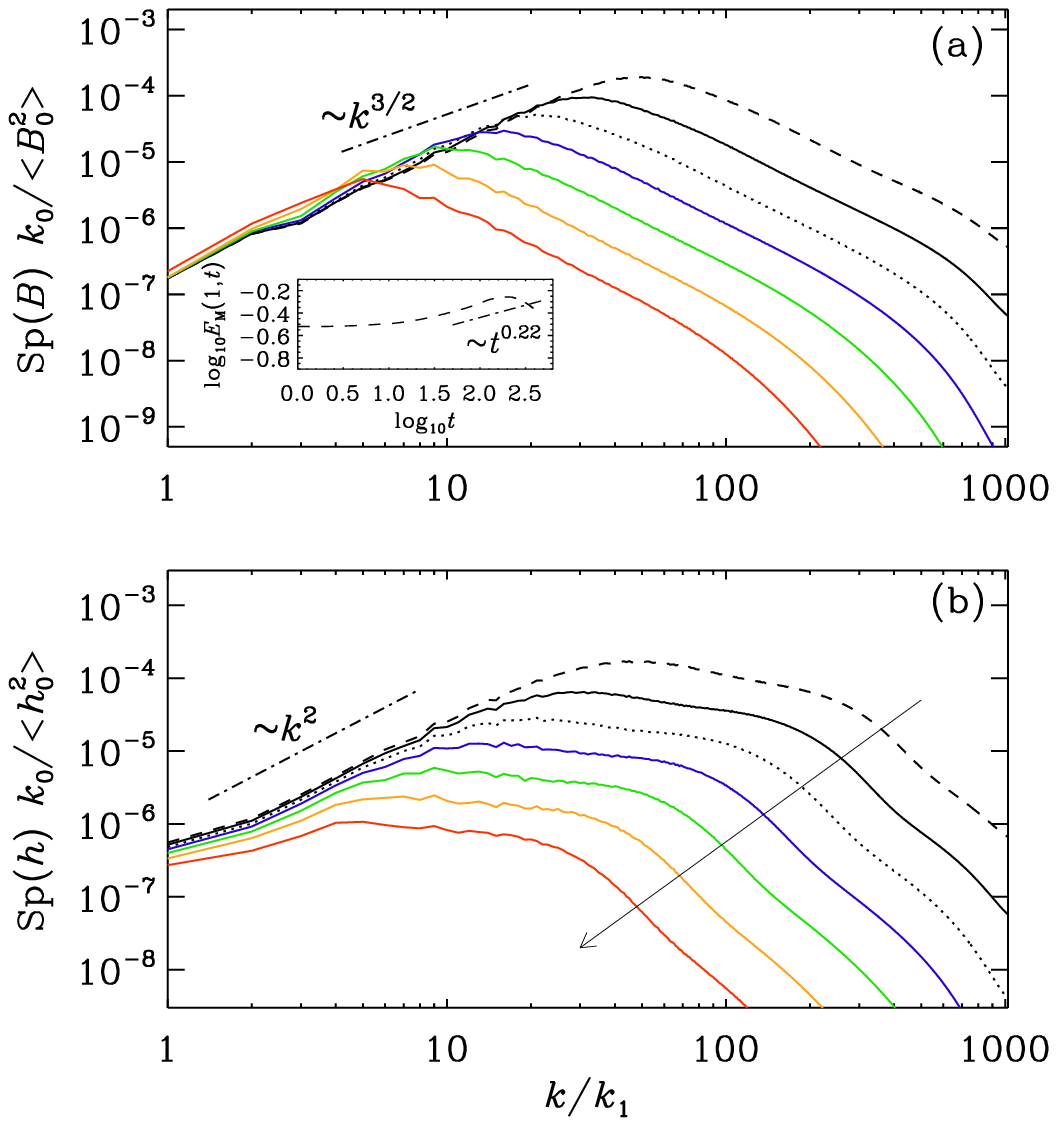}
\end{center}\caption[]{
\blue{
Similar to \Fig{rspec_select_hoskM_k60del2bc_k3}, but for $\alpha=2$
and $k_0=180\,k_1$ at $t=2$, 6, 15, 34, 80, 183, and 416.
The inset applies here to the evolution at $k=5\,k_1$,
instead of $k=k_1$, as for all other plots.
}
}\label{rspec_select_hoskM_k180del2bc_k2p0a2}\end{figure}

\blue{
A large scale-separation ratio, $k_0/k_1$, was previously found to
be important.
For example, in the context of the Hall cascade, a three-fold larger
value of $k_0/k_1$ was needed to demonstrate clear evidence for inverse
cascading \citep{Bran20}.
Therefore, we now present in \Fig{rspec_select_hoskM_k180del2bc_k2p0a2}
the results for $k_0=180\,k_1$.
We see that, similarly to the case of $\alpha=1.7$ in the inset of
\Figp{rspec_select_hoskM_k60del2bc_k1p7}{a}, there is an initial rise
of spectral magnetic energy compatible with being $\propto t^{0.22}$,
which, again, is followed by a decline at very late times.
This result therefore supports the notion that the Hosking integral is
indeed well conserved and that it governs the evolution of the magnetic
field even for $\alpha=2$.
}

\subsection{Evolution in the $pq$ diagram}

There is a range of tools for assessing the decay properties of MHD
turbulence.
We did already discuss the determination of $I_{\rm H}$ and $I_{\rm SM}$,
and the potentially universal coefficients $C_{\rm H}^{(\xi)}$,
$C_{\rm H}^{({\cal E})}$, and $C_{\rm H}^{({E})}$.
We also discussed the close relation between the envelope parameter
$\beta$ in \Eqs{Compensated}{GeneralFits}, and the parameter $q$
characterizing the growth of the correlation length $\xiM\propto t^q$.
There is also the parameter $p$ characterizing the decay of magnetic
energy, $\EEM\propto t^{-p}$.
Both $p$ and $q$ can also be determined as instantaneous
scaling parameters through $p(t)=-\dd\ln\EEM/\dd\ln t$ and
$q(t)=\dd\ln\xiM/\dd\ln t$, and their parametric representation $p(t)$
versus $q(t)$ gives insights about the properties of the system and how
far it is from a self-similar evolution \citep{BK17} and the
scale-invariance line, $p=2(1-q)$; see \cite{Zhou+22}.

%FIG7
\begin{figure}\begin{center}
\includegraphics[width=.7\columnwidth]{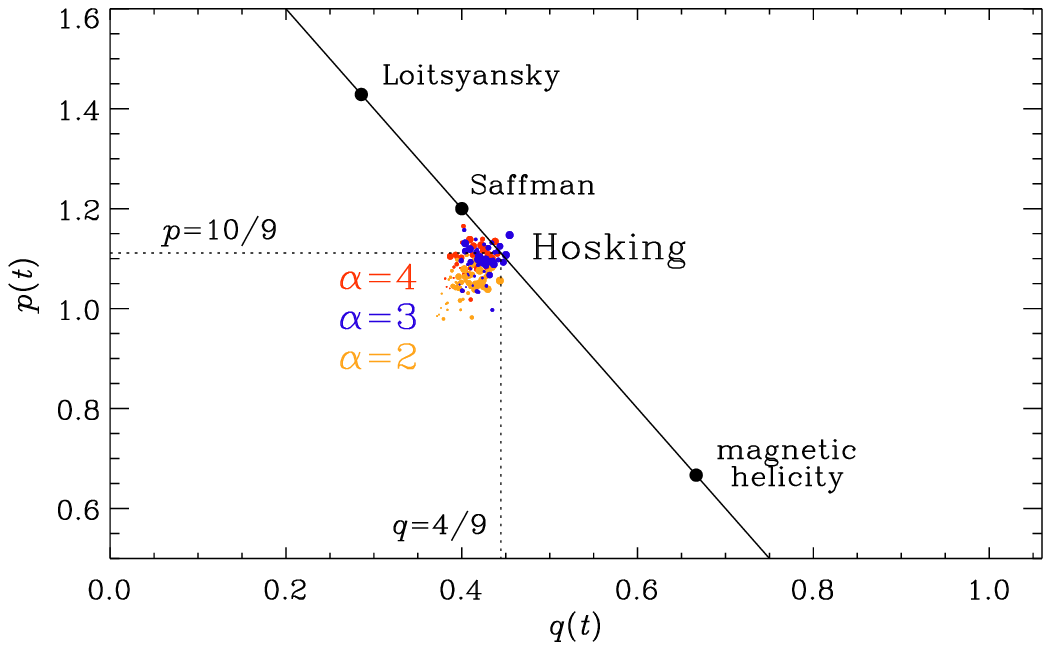}
\end{center}\caption[]{
$pq$ diagram showing a parametric representation of $p(t)$ vs $q(t)$
for Runs~B ($\alpha=3$, blue), C ($\alpha=4$, red), and Q ($\alpha=2$, orange)
and $10<t<60$.
Larger symbols correspond to later times.
The locations for Loitsyansky and Saffman scalings, as well as for
the fully helical case are indicated as black dots along
the scale-invariance line (black solid line),
$p=2(1-q)$, and the black dotted lines mark the position
$q=4/9$ and $p=10/9$.
}\label{pEMxi_pq_run_3}
\end{figure}

In \Fig{pEMxi_pq_run_3}, we show such a $pq$ diagram for Runs~B, C, and Q.
We see that the points $(q,p)$ for different times and for both runs
cluster around $(q,p)=(4/9,\,10/9)$, as expected for Hosking scaling.
The locations for Loitsyansky and Saffman scalings, $(2/7,\,10/7)$
and $(2/5,\,6/5)$, respectively, as well as for the fully helical
case $(2/3,\,2/3)$ are also indicated for comparison.
Note that, even for Run~Q with $\alpha=2$, the points are closer to
Hosking scaling than to Saffman scaling.

A detailed assessment of the full range of scaling parameters is important
for establishing the validity of Hosking scaling.
Assessments based on comparisons of the parameter $p$ for different
runs may not be sufficient, and have led to inconclusive results;
see \cite{Armua+23} for recent results.
Thus, the idea behind the Hosking phenomenology is therefore not
universally accepted.
Possible reasons for discrepancies could lie in an insufficiently large
magnetic Reynolds number and therefore also in a lack of a sufficiently
long inertial range.
Therefore, it would be useful to have independent verification from
other groups.
In this connection, it should be noted that additional support for the
validity of Hosking scaling came from two rather different numerical
experiments.
First, in applications to the Hall cascade, the Hosking phenomenology
predicts the scalings $q=4/13$ and $p=10/13$, which was confirmed
by simulations \citep{Bra23}.
Second, in relativistic plasmas where the mean magnetic helicity
density is finite, but the total chirality vanishes because the helicity
is exactly balanced by fermions chirality, the Hosking phenomenology
predicts a decay of mean magnetic helicity $\propto t^{-2/3}$, which,
again, was confirmed by simulations \citep{BKS23}.

\section{Conclusions}
\label{Conclusions}

Our work has shown that the decay dynamics of an initial magnetic field
with power law scaling proportional to $k^3$ is similar to that for $k^4$.
According to a simple argument involving self-similarity, we showed
and confirmed that the temporal growth of the magnetic energy spectra
at small $k$ is proportional to $t^{4\alpha/9-2/3}$, so for $\alpha=3$,
we have an increase proportional to $t^{2/3}$, while for $\alpha=4$,
the increase is proportional to $t^{10/9}$.
Thus, although we cannot exclude the possibility of artifacts from
the finite size of the computational domain, our simulations now
suggest inverse cascading even for an initial Saffman spectrum.
This underlines the importance of the Hosking integral in determining
the decay dynamics for a large class of initial magnetic energy spectra.
We also confirmed that the nondimensional coefficients in the
empirical scaling relations for $\xiM(t)$, $\EEM(t)$, and $\EM(k,t)$
are compatible with those found earlier for an initial $k^4$ subinertial
range spectrum.

At the moment, even with a resolution of $2048^3$ mesh points, we
cannot make very firm statements about the case $\alpha=1.7$, because
$I_{\rm H}$ is not sufficiently well conserved and the value of $\alpha$
is close to 3/2.
It would be useful to reconsider the case $\alpha=2$ with even higher
resolution to confirm the violation of the conservation of the magnetic
Saffman integral, and thus weak inverse cascading $\propto t^{0.2}$.

\section*{Acknowledgements}
We are grateful to David Hosking, Antonino Midiri, Alberto Roper Pol,
and Kandaswamy Subramanian for encouraging discussions.
\blue{
We are also grateful to Hongzhe Zhou for his suggestion to consider
steeper spectra, as we have now discussed in \App{Steeper}.
}
We thank the two referees for useful suggestions and comments
that have led to improvements of the paper.
RS acknowledges the funding support provided by the ERC HERO-810451 grant.
TV thanks CERN for hospitality during the course of this work. 

\section*{Funding}
A.B.\ and R.S.\ where supported in part by the Swedish Research Council
(Vetenskapsr{\aa}det, 2019-04234); Nordita is sponsored by Nordforsk.
T.V.\ was supported by the U.S. Department of Energy, Office of High
Energy Physics, under Award No.~DE-SC0019470.
We acknowledge the allocation of computing resources provided by the
Swedish National Allocations Committee at the Center for Parallel
Computers at the Royal Institute of Technology in Stockholm and
Link\"oping.

\section*{Declaration of Interests}
The authors report no conflict of interest.

\section*{Data availability statement}
The data that support the findings of this study are openly available
on Zenodo at doi:10.5281/zenodo.8128611 (v2023.07.09).
All calculations have been performed with the {\sc Pencil Code}
\citep{JOSS}; DOI:10.5281/zenodo.3961647.

\section*{Authors' ORCIDs}

\noindent
A. Brandenburg, https://orcid.org/0000-0002-7304-021X\\
R. Sharma,      https://orcid.org/0000-0002-2549-6861\\
T. Vachaspati,  https://orcid.org/0000-0002-3017-9422 	

\appendix

\section{\blue{Approach to a $k^4$ spectrum from a steeper one}}
\label{Steeper}

\begin{figure}\begin{center}
\includegraphics[width=.62\columnwidth]{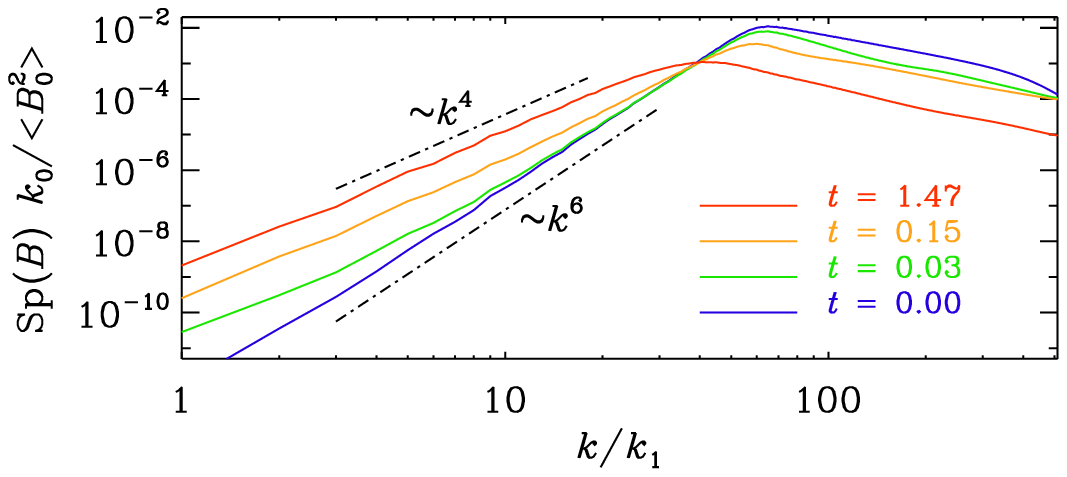}
\end{center}\caption[]{
\blue{
Approach to a $k^4$ spectrum from a steeper $k^6$ spectrum for
$k_0=60\,k_1$ using $1024^3$ mesh points and otherwise the same
parameters as for \Fig{rspec_select_hoskM_k60del2bc_k3}.
}
}\label{rspec_select_k60del2bc_k6b}\end{figure}

\blue{
In this paper, we focus on the case $\alpha<4$.
This is because for $\alpha=4$, the spectrum quickly develops
into one that is equivalent to $\alpha=4$.
The approach to a $k^4$ spectrum from a steeper $k^6$ spectrum
is shown in \Fig{rspec_select_k60del2bc_k6b}.
We see that the spectra quickly gain power at low $k$ so that the
subinertial range is $\propto k^4$.
This happens at very early times, well before any inverse cascading
has started yet.
}

%FIG8

\section{Finite-size effects}
\label{FiniteSizeEffects}

\begin{figure}\begin{center}
\includegraphics[width=.62\columnwidth]{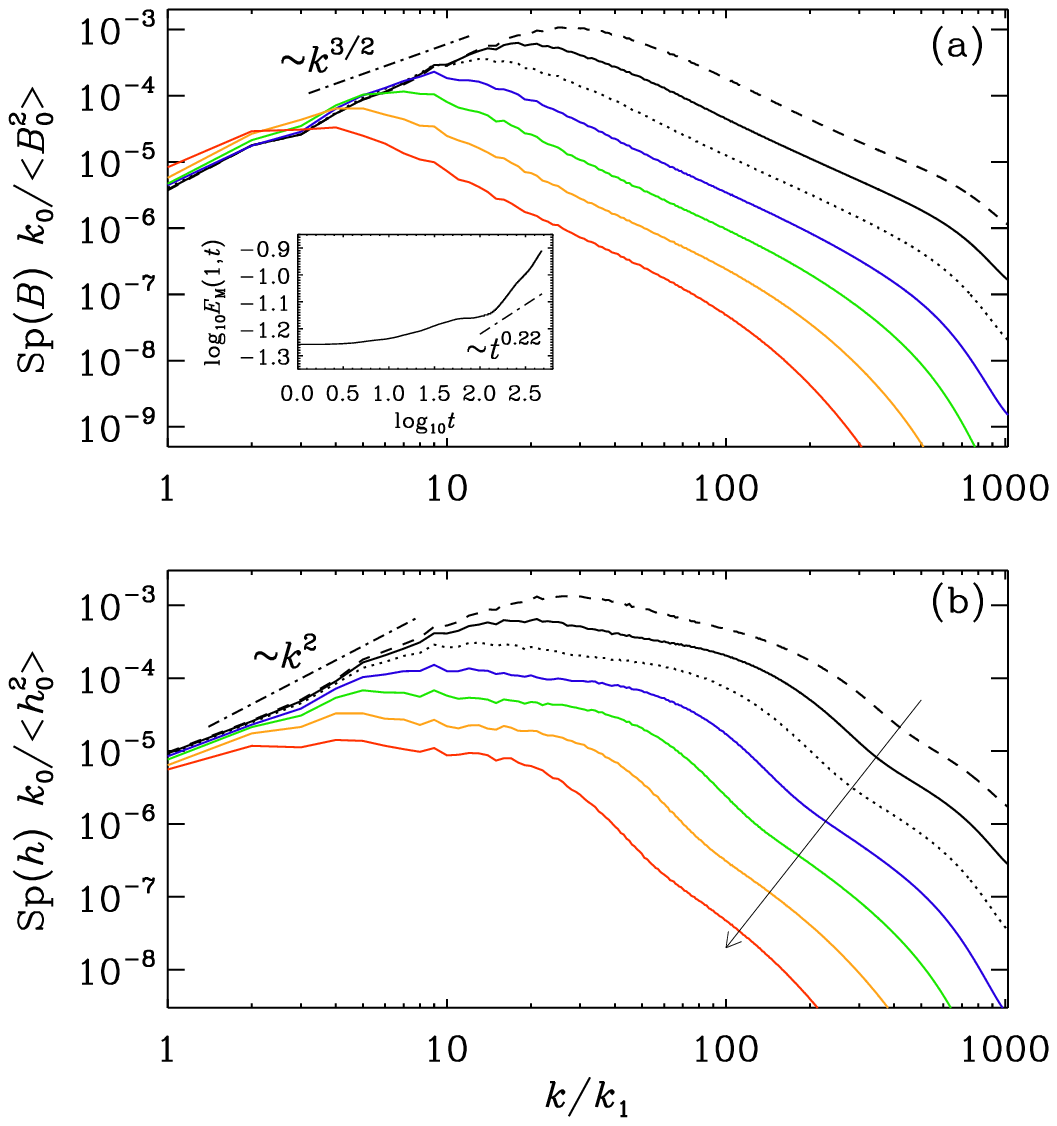}
\end{center}\caption[]{
\blue{
Similar to \Fig{rspec_select_hoskM_k60del2bc_k3}, but for $\alpha=2$
and at $t=2$, 6, 15, 37, 87, 205, and 475 and with $k_0=60\,k_1$.
}
}\label{rspec_select_hoskM_k60del2bc_k2p0}\end{figure}

\blue{
In \Sec{SaffmanCase}, we mentioned that we needed a larger
scale-separation ratio to obtain reliable results for $\alpha=2$.
To demonstrate the problem, we show here the result for the usual
scale separation of $k_0/k_1=60$.
The inset to \Figp{rspec_select_hoskM_k60del2bc_k2p0}{a} shows that
the growth of $\EM(k,t)$ at $k=k_1$ does not follow clear power law scaling.
There is a decline in the slope in the range $50<t<100$, followed by an
increase that lasts until the end of the simulation at $t=475$.
A likely explanation for this unexpected behavior could be finite
size effects.
If that is the case, the intermediate decline in the slope could be
interpreted as evidence for a levelling off, compatible with Saffman
scaling.
}

\blue{
We should also mention that it turned out that, even for $k_0=60\,k_1$,
we had to decrease the initial magnetic field strength to
$\vAz/\cs\approx0.65$ to prevent the code from crashing.
This value of $\vAz/\cs$ is about 30\% smaller than our usual
value of $\vAz/\cs\approx0.87$ that was used
for the other runs at that resolution.
While these field strengths are not that different, it indicates that
at early times, our simulations are close to the limit below which we
can still trust them.
}

%FIG9

\bibliographystyle{jpp}
\bibliography{ref}
\end{document}